\documentclass[]{emulateapj}

\slugcomment{Submitted to  Astrophysical Journal Letters}

\shorttitle{Axisymmetric Final Parsec Problem}
\shortauthors{Fazeel M. Khan \& Kelly Holley-Bockelmann}

\begin{document}

\title{Supermassive Black Hole Binary Evolution in Axisymmetric Galaxies: the final parsec problem is not a problem}

\author{Fazeel Mahmood Khan\altaffilmark{1}}
\author{Kelly Holley-Bockelmann\altaffilmark{2, 3}}
\affil{$^1$Department of Physics, Government College University (GCU), 54000  
Lahore, Pakistan; khan@ari.uni-heidelberg.de}
\affil{$^2$Department of Physics and Astronomy, Vanderbilt University, Nashville, TN, 37235; k.holley@vanderbilt.edu}
\affil{$^3$Department of Physics, Fisk University, Nashville, TN, 37208}

\begin{abstract}

During a galaxy merger, the supermassive black hole (SMBH) in each galaxy is thought to sink to the center of the potential and form a supermassive black hole binary; this 
binary can eject stars via 3-body scattering, bringing the SMBHs ever closer. In a static
spherical galaxy model, the binary stalls at a separation of about a parsec after ejecting all the stars in its loss cone -- this is the well-known {\it final parsec problem.}
Earlier work has shown that the centrophilic orbits in triaxial galaxy models are key in refilling the loss cone at a high enough rate to prevent the black holes from stalling. 
However, the evolution of binary SMBHs has never been explored in axisymmetric galaxies, so it is not clear if the final parsec problem persists in these systems. 
Here we use a suite of direct $N$-body simulations to follow SMBH binary evolution in galaxy models with a range of ellipticity. For the
first time, we show that mere axisymmetry can solve the final parsec problem; we find the the SMBH evolution is independent of $N$ for an axis ratio of $c/a=0.8$, and that
the SMBH binary separation reaches the gravitational radiation regime for $c/a=0.75$.

\end{abstract}


\section{Introduction}\label{sec-intro}

Supermassive Black Holes (SMBHs), with masses from $10^6-10^{10} M_\odot$ are thought to dwell at the centers of most galaxies~\citep[e.g.][]{kr95,FF05}, and since we know that 
galaxy mergers are commonplace, the formation of a binary SMBH is almost inevitable at some point during galaxy assembly~\citep{BBR80}. There are cases in which there is a clear 
observational evidence for two widely separated SMBHs\citep[e.g.][]{comerford09}, as well as growing evidence for true SMBH binaries \citep{komossa06,bon12}. 
Upon coalescence, SMBH binaries would be the highest signal-to-noise ratio sources of gravitational waves, detectable from essentially 
any cosmological redshift, for future space-borne gravitational wave detectors such as the {\it New Gravitational wave Observatory} 
(NGO) \citep{Hughes03,BC04,LISA2011}.  
 
The paradigm for SMBH binary evolution consists of three distinct phases \citep{BBR80}. 
First, the two SMBHs sink towards the galaxy center due to the dynamical friction, eventually forming a bound pair with 
semi-major axis $a \sim r_\mathrm{h}$, where $r_\mathrm{h}$ is 
the binary's influence radius, typically is defined as the radius enclosing twice the mass of the binary in 
stars $M(<r_\mathrm{h})=2 M_\bullet$, where $M_\bullet$ is the mass of the binary. Dynamical friction continues to drive the SMBHs closer until the system 
becomes a {\it hard binary}; here the separation is $a \sim a_\mathrm{h}:$ \citep{Q96,Yu02}
\begin{equation}
a_\mathrm{h} := \frac{G \mu_r}{4 \sigma^2},
\label{eqn1}
\end{equation}
where $\mu_\mathrm{r}$ is the binary's reduced mass and $\sigma$ is the $1$D velocity dispersion. In this second phase, 
{\it slingshot} ejection of stars is the dominant mechanism for taking energy and angular momentum away from 
SMBH binary. If 3-body scattering can shrink the binary orbit by a factor of order a hundred, then the SMBHs will be close enough to emit copious gravitational radiation.  
This gravitational wave emission in the third and final phase will drain orbital energy 
from the binary and will drive the final coalescence in roughly 100 Myr. The transition from the first to the 
second phase can be quite prompt, provided that the galaxy mass ratio is not too small ($q \gtrsim 0.1$) 
\citep{calleg11}. In contrast, the subsequent
transition from the 3-body scattering to gravitational radiation regime could constitute a bottleneck for the binary
coalescence, simply caused by a lack of stars which can interact with the binary. This is
the so-called {\it Final Parsec Problem}.

Theorists commonly invoke gas to move these SMBHs through the final parsec. This is because
the SMBHs can frict against the gas very efficiently, as long as the gas is not forming 
stars~\citep[e.g][]{mayer07,escala05}. Unfortunately, relying on gas to solve the Final Parsec
Problem will leave out the most massive SMBHs, since their host galaxies are massive ellipticals
and are therefore gas-poor~\citep{kannappan04}.

There has been significant work to describe the Final Parsec Problem and to solve it with a purely stellar dynamical approach. For example, \cite{Yu02} first pointed out, based on an analysis of an HST sample of nearby elliptical galaxies and spiral 
bulges from \cite{Faber97}, that flattening and non-axisymmetry could bring SMBH binaries 
to coalescence. Later, several studies built self-consistent cuspy, triaxial models with a single central SMBH and showed 
that models could be concocted with a significant fraction of centrophilic orbits that would efficiently drive 
the hardening rate of a binary if it were present~\citep[e.g.][]{MP2004, khb02, khb06}.
Also direct $N$-body simulations of SMBH binary evolution in isolated galaxy models showed that the Final Parsec Problem could potentially be solved by purely stellar dynamical models that develop a significant amount of triaxiality~\citep{ber06,ber09}.  
Recent numerical studies showed that galaxies which form via mergers are mildly triaxial \citep{kh11,Preto11,kh12}. Hardening 
rates of SMBH binaries resulting from equal-mass mergers are substantially higher than those found in spherical nuclei, and are independent of the total number $N$ of stars.
Presumably, the triaxial shapes of merger remnants reported in these studies result in a large population of stars 
on regular centrophilic orbits deep within the galaxy potential, such as pyramid or box orbits with formally zero angular momentum~\citep{gb85,Valluri98,khb01}. Outside the influence radius of the SMBH, 
many of the centrophilic orbits are chaotic in a triaxial potential, though some
regular resonant boxlet orbits remain~\citep{pm01,khb02}. 

In contrast to the wealth of work on solving the final parsec problem in triaxial galaxies, comparatively little attention has been paid to exploring
binary coalescence in axisymmetric systems. Recent work on the capture of stars from the loss cone in axisymmetric nuclei looks promising for the final parsec problem, in that the stellar capture rate is several times higher in flattened systems~\citep{Magorrian99,vm13}. In an axisymmetric galaxy, the centrophilic orbit family includes saucer or cone orbits~\citep{st99}, and these
can, in principle increase 3-body scattering rates. However, no detailed N-body simulations have been done to test this. 
In this paper, we explore SMBH coalescence in axisymmetric galaxies using high resolution N-body simulations of equilibrium axisymmetric galaxy models with a range of intrinsic flattening.

\section{Initial conditions and numerical methods}\label{sec-model}

\subsection{The host galaxies and their SMBHs}

We generated our equilibrium axisymmetric systems from initially spherical SMBH-embedded $\eta$ models~\citep{Tremaine94}
by applying a slow, steady velocity drag on the particles in the z direction~\citep[see][for details]{khb01}. This {\it adiabatic squeezing} technique 
preserves the phase space distribution function of the system, so although the orbit content changes, the density profile remains unchanged 
as the galaxy shape slowly evolves~\citep{khb02}. The suite we present here has an inner density slope, $\gamma$, of 1.0.  Our model has
a mass of 1.0 in system units, the central SMBH has a mass of 0.005, and the half mass radius is 2.41; scaling the SMBH mass to the Milky Way implies that we
are simulating the inner few hundred parsecs of the galaxy.
A second equal mass SMBH is introduced into the outskirts of the equilibrium model at a distance of 0.5 in model units with 70 $\%$ of the circular velocity.
  
\begin{table}
\caption{Parameters of the SMBH binary study} 
\centering
\begin{tabular}{c c c c c c c c c}
\hline
Run & $N$ & $\gamma$ & $c/a$ & $q$\\
\hline
Spha & 1000k & $1.0$& $1.0$& $1.0$\\
Flat9a & 1000k & $1.0$& $0.9$& $1.0$\\
Flat9b & 800k & $1.0$& $0.9$& $1.0$\\
Flat9c & 500k & $1.0$& $0.9$& $1.0$\\
Flat9d & 250k & $1.0$& $0.9$& $1.0$\\
Flat9e & 125k & $1.0$& $0.9$& $1.0$\\[0.2ex]
Flat8a & 1000k &	$1.0$& $0.8$& $1.0$\\
Flat8b & 800k &	$1.0$& $0.8$& $1.0$\\
Flat8c & 500k & $1.0$& $0.8$& $1.0$\\
Flat8d & 250k & $1.0$& $0.8$& $1.0$\\
Flat8e & 125k & $1.0$& $0.8$& $1.0$\\[0.2ex]
Flat75a & 1000k & $1.0$& $0.75$& $1.0$\\
Flat75b & 800k & $1.0$& $0.75$& $1.0$\\
Flat75c & 500k & $1.0$& $0.75$& $1.0$\\
Flat75d & 250k & $1.0$& $0.75$& $1.0$\\
Flat75e & 125k & $1.0$& $0.75$& $1.0$\\[0.2ex]

\hline
\end{tabular}\label{TableA}
\tablecomments{Column 1: Galaxy model. Column 2: Number of particles. Column 3: Central density slope $\gamma$.
 Column 4: Axes ratio. Column 5: SMBH mass ratio.}
\end{table}

\subsection{Numerical Methods and Hardware}

The $N$-body integrations are carried out using the $\phi$GRAPE\footnote{\tt
ftp://ftp.ari.uni-heidelberg.de/staff/berczik/phi-GRAPE/} \citep{harfst}, a parallel,
direct-summation $N$-body code that uses Graphic Processing
Units (GPU) cards to accelerate the computation of pairwise gravitational forces between all
particles.
$\phi$GRAPE code
uses a 4th-order Hermite integration scheme to advance all
particles with hierarchical individual block time-steps,
together with the parallel usage of GPU cards to calculate acceleration ${\vec  a}$ and its first time derivative ${\dot{{\vec a}}}$ 
between all particles. For the force calculation we use 
the {\tt SAPPORO} \citep{gab09} library which emulates standard
GRAPE-6 library calls on the GPU.

As { $\phi$GRAPE} does not include the
regularization \citep{mikkola98} of close encounters or binaries,
we use a gravitational Plummer-softening for force calculations. The
softening length for star-star encounters is chosen to be 
smaller (i.e. $ 10^{-4}$ in our model units) than the minimum separation reached by SMBH binaries in our study. The softening between the SMBH particles is set equal to zero.

The suite of  $N$-body experiments were carried out on a high-performance GPU computing cluster {\tt ACCRE} employing 192 GPUs at Vanderbilt University, Nashville, TN. 

For a detailed discussion on the $N$-body methods and choice of 
parameters, we refer the reader to~\cite{kh11}. 

\section{SMBH BINARY EVOLUTION IN FLATTENED GALAXY MODELS}\label{Resutls}

We created models with axis ratios $c/a = [1.0,0.9,0.8,0.75]$ (see table \ref{TableA} ) in order to study the SMBH binary evolution as a function of flatness.

\begin{figure}
\centerline{
  \resizebox{0.85\hsize}{!}{\includegraphics[angle=270]{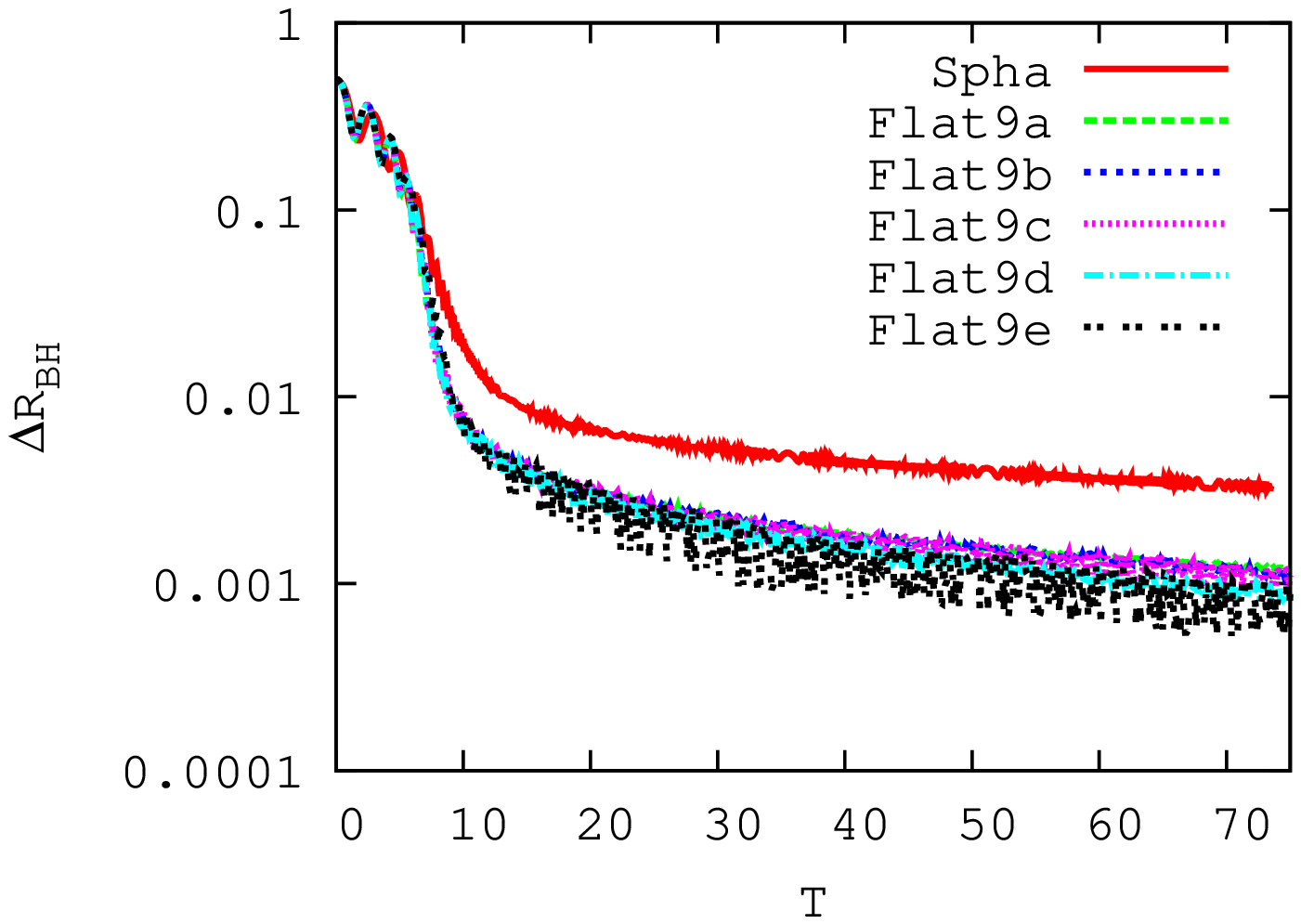}}
  }
\centerline{
  \resizebox{0.85\hsize}{!}{\includegraphics[angle=270]{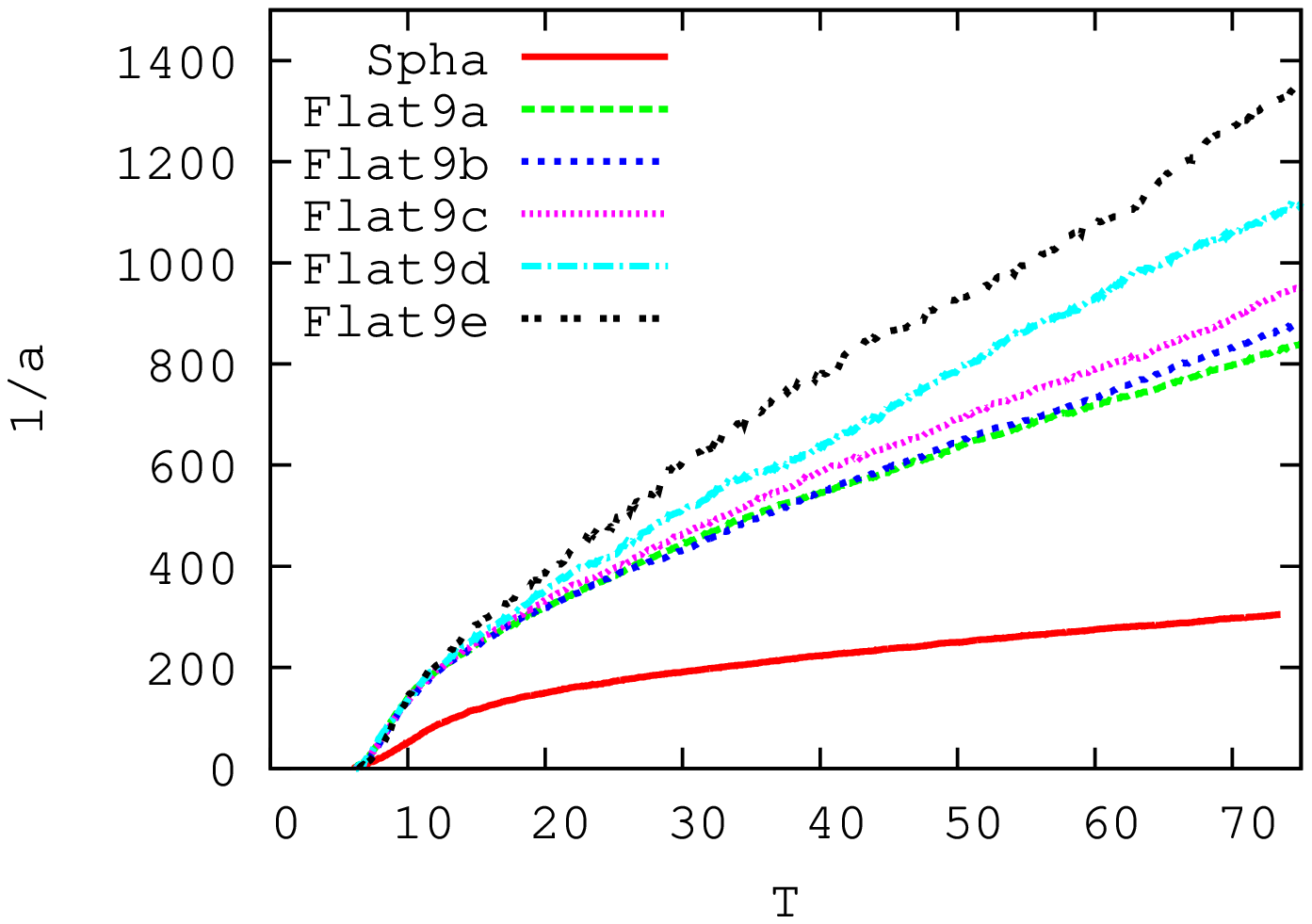}}
  }
  \centerline{
  \resizebox{0.85\hsize}{!}{\includegraphics[angle=270]{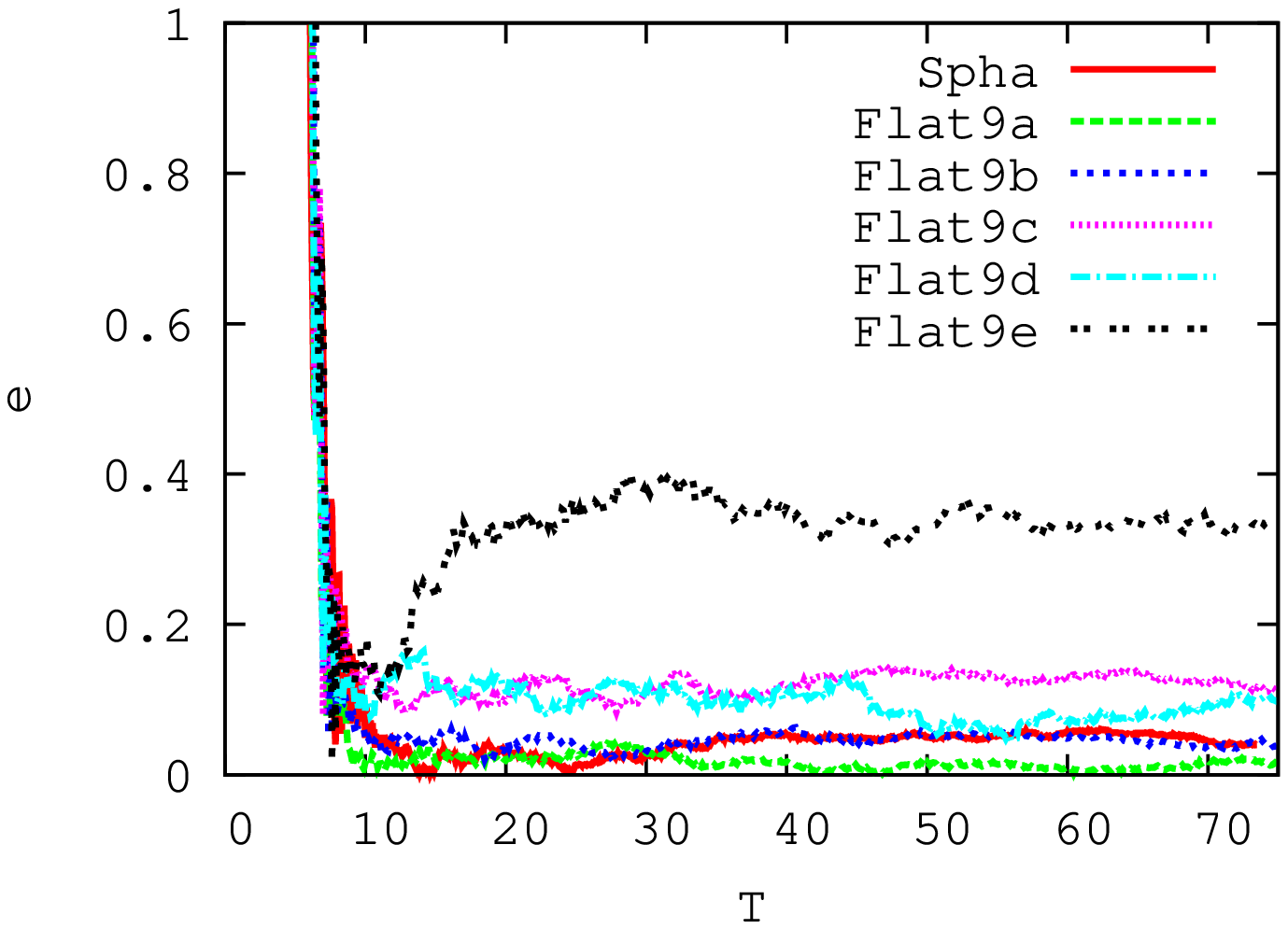}}
  }
\caption[]{
Evolution of the separation between each SMBH (top), the semimajor axis (middle) and eccentricity (bottom) of the SMBH binary in $N$-body integrations for spherical and flattened models with $c/a$ = 0.9 (see table \ref{TableA}). Time and $R$ are measured in model units. 
} \label{flat1}
\end{figure}

Figure \ref{flat1} describes the comparison between the evolution of the SMBH binary in spherical and flattened galaxy models ($c/a = 0.9$). In the first phase,  which can be seen before $T=10$,  the separation between the two SMBHs shrink due to dynamical friction. The evolution of SMBHs in this phase is very similar for both the spherical and flattened galaxy models. However the subsequent evolution, governed by three body encounters, happens at much faster rate in flat galaxy models, as can be seen from the evolution of both the separation and the semimajor axis of the binary SMBH (the top and middle panels). The SMBH binary forms with an eccentricity, $e$,  much less than 0.1 and it remains small during the binary evolution. Although the semi-major axis of SMBH binary evolves faster than that of spherical galaxy model, it still shows clear dependence on N (see middle panel) -- a sure sign that the binary evolution in this simulation is plagued by the numerical effects of 2-body relaxation. 

The binary hardening rates $s$ are calculated by fitting straight lines to $a^{-1}(t)$ in the late phase of binary evolution from $T = 30 - 60$. The value of $s$ decreases with increasing $N$, and again this  suggests that numerical 2-body relaxation is important for SMBH binary evolution in both of these models. Note, though,  that $s$ is roughly four times higher for the SMBH binary evolution in flattened system ($c/a = 0.9$) compared to a spherical galaxy model. In flattened galaxy models, then,  there appears to be an additional supply of stars that interact with the massive binary.

\begin{figure}
\centerline{
  \resizebox{0.98\hsize}{!}{\includegraphics[angle=270]{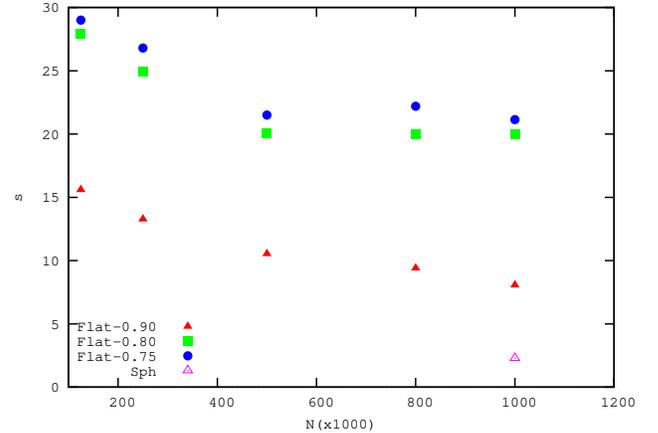}}
  }
\caption[]{
Average hardening rates of SMBH binary in spherical and flattened galaxy models    
in Table \ref{TableA}. The average is measured between $T=40$ and $T=70$ in model
units.  
} \label{hrates}
\end{figure}

\begin{figure}
\centerline{
  \resizebox{0.85\hsize}{!}{\includegraphics[angle=270]{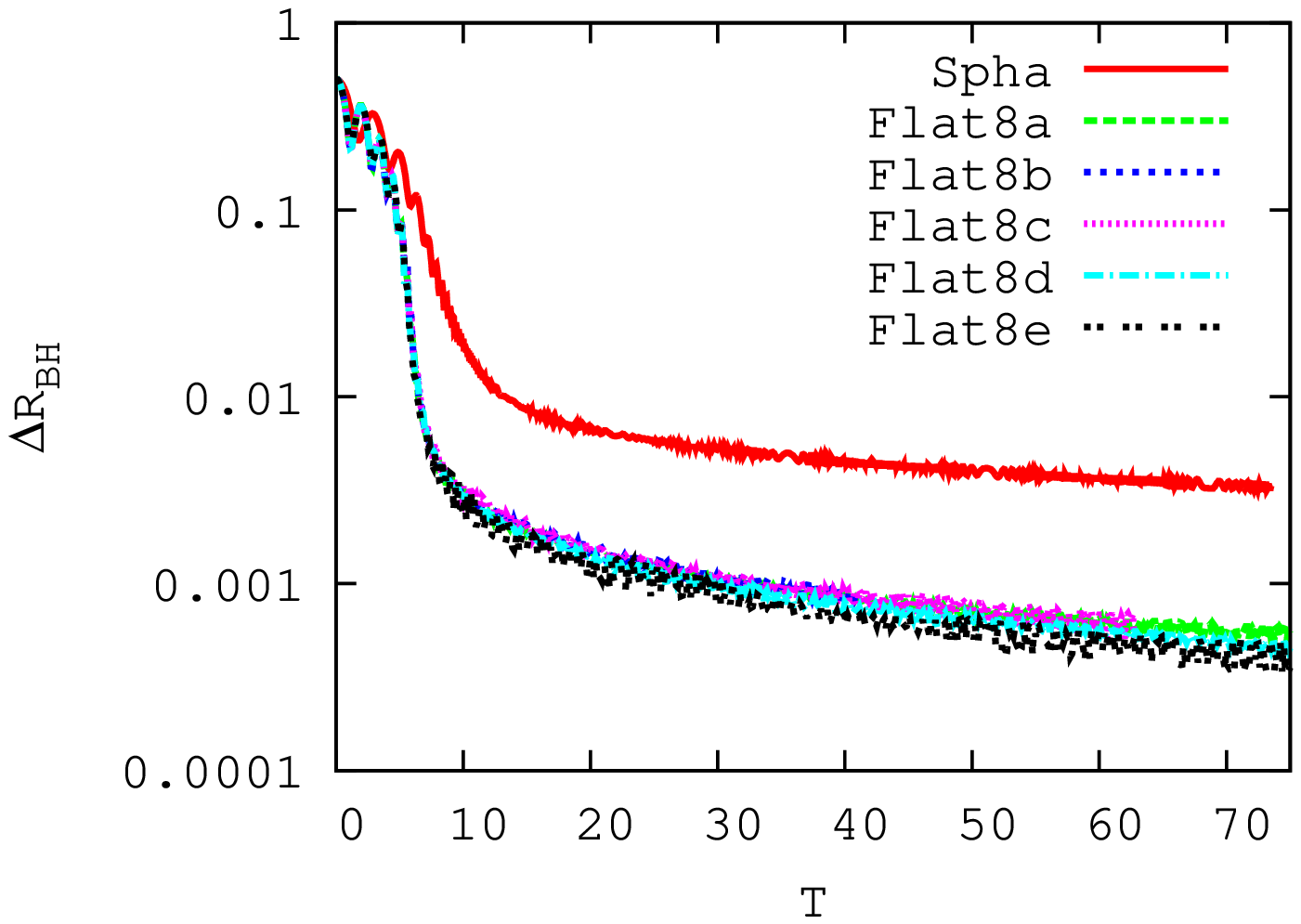}}
  }
\centerline{
  \resizebox{0.85\hsize}{!}{\includegraphics[angle=270]{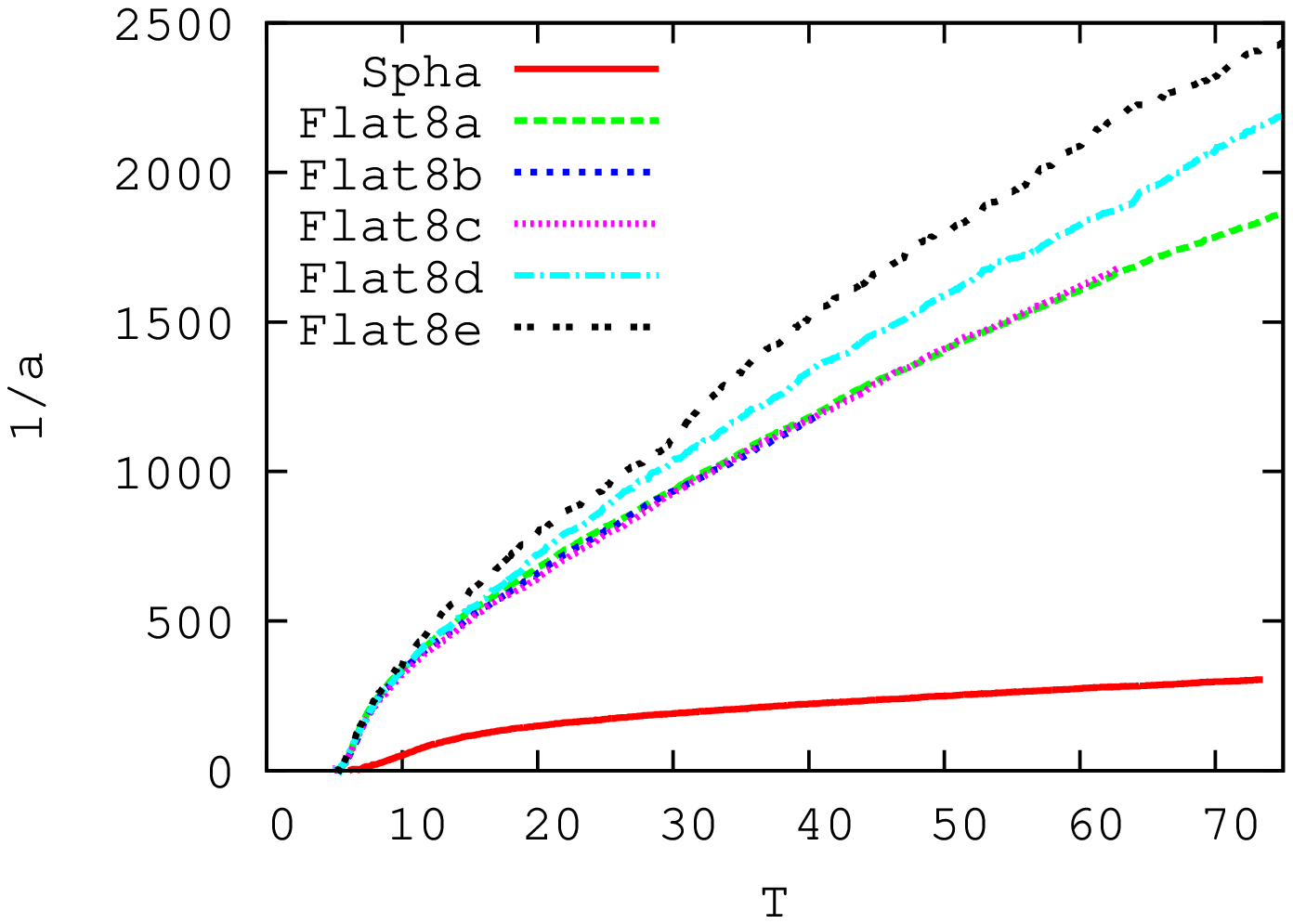}}
  }
  \centerline{
  \resizebox{0.85\hsize}{!}{\includegraphics[angle=270]{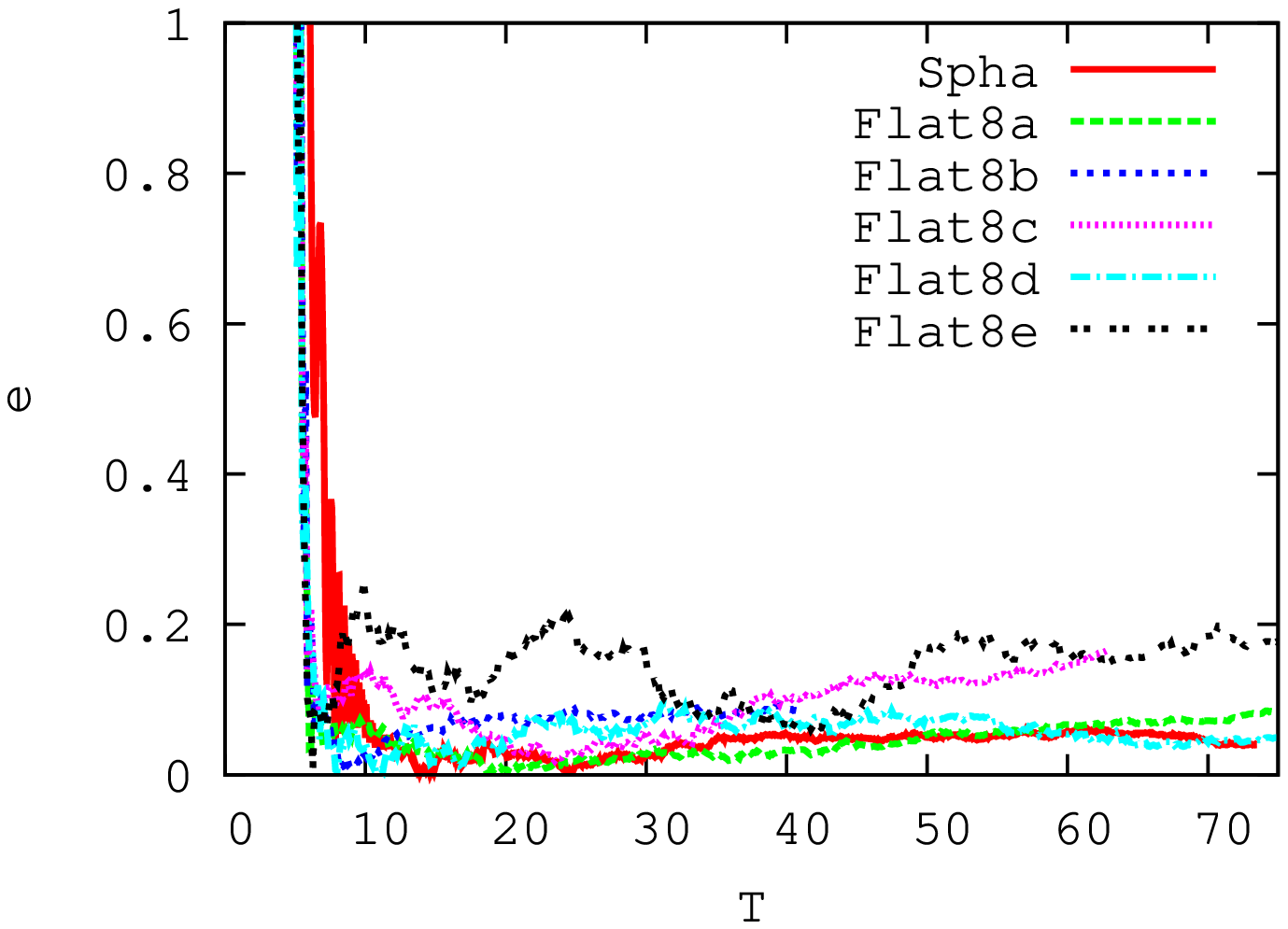}}
  }
\caption[]{
Evolution of the binary SMBH separation (top), semimajor axis (middle) and eccentricity (bottom) of the SMBH binary in $N$-body integrations for spherical and flattened models with $c/a = 0.8$  (see table \ref{TableA}). Time and $R$ are measured in model units.
} \label{flat2}
\end{figure}

The SMBH binary evolution for non-rotating models with $c/a = 0.8$ are shown in figure \ref{flat2}. On the face of it, the evolution of SMBH binary in flattened models with $c/a = 0.8$ is very similar to $c/a = 0.9$. However, there is a striking difference for runs with N greater than 500k in the sense that SMBH binary evolution becomes independent of N. This implies that the system is no longer dominated by artificial 2-body relaxation, and we are uncovering the accurate SMBH binary evolution. Again, we calculated SMBH binary hardening rates in interval between $T = 30$ and $T = 60$. Note that for $N>500$k, $s=22$ is 8 times higher than the spherical galaxy model at the highest particle resolution.

\begin{figure}
\centerline{
  \resizebox{0.85\hsize}{!}{\includegraphics[angle=270]{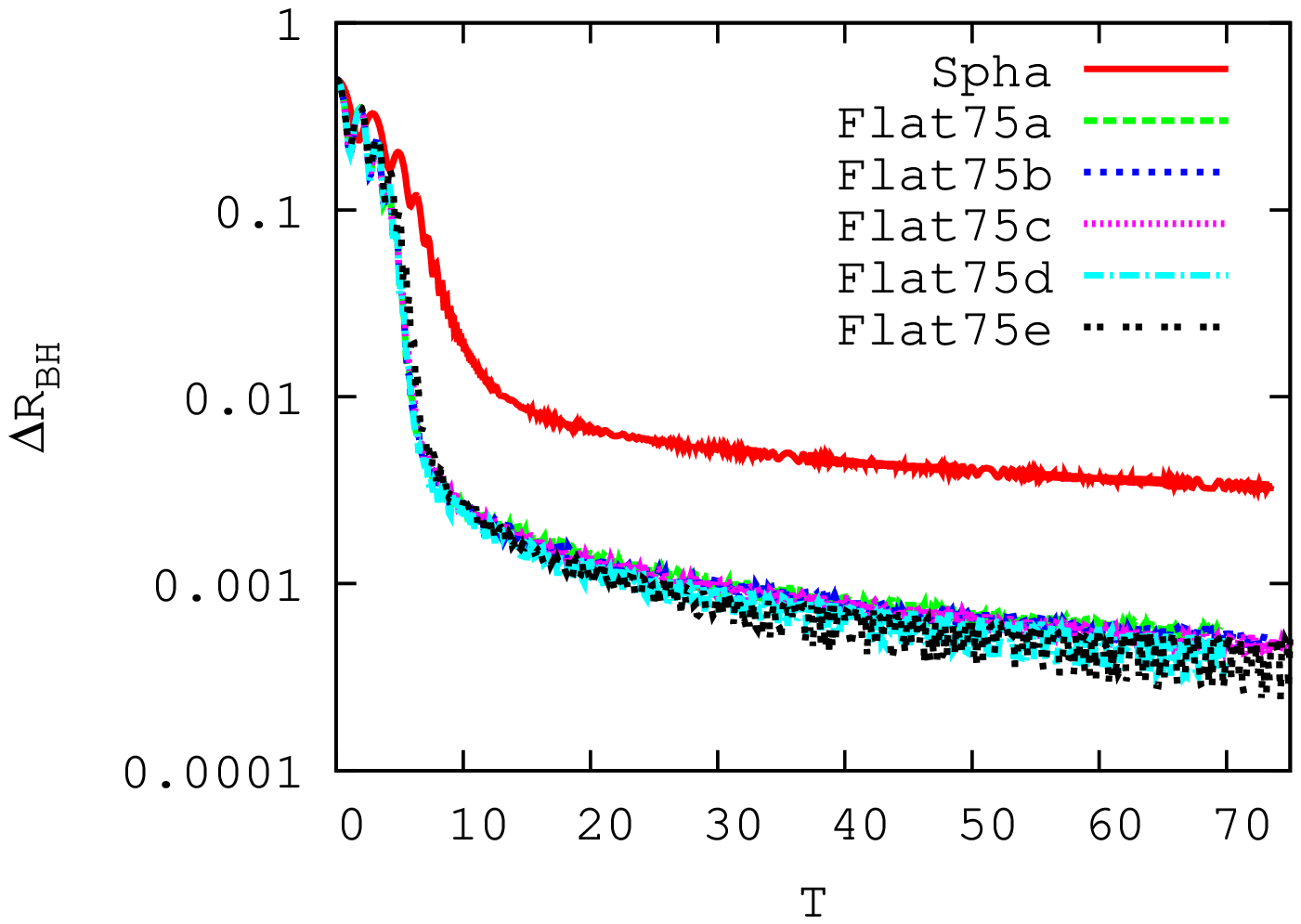}}
  }
\centerline{
  \resizebox{0.85\hsize}{!}{\includegraphics[angle=270]{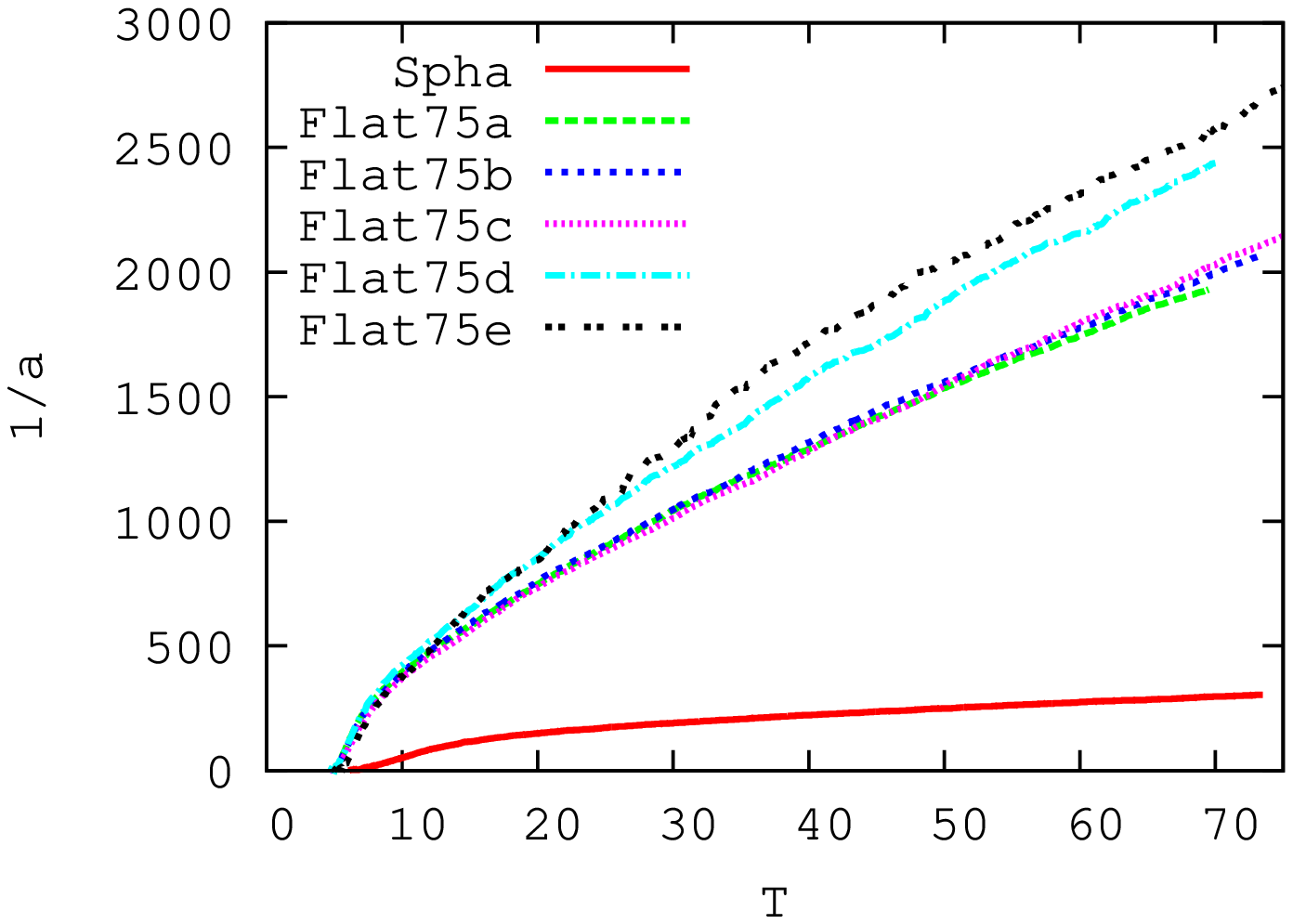}}
  }
  \centerline{
  \resizebox{0.85\hsize}{!}{\includegraphics[angle=270]{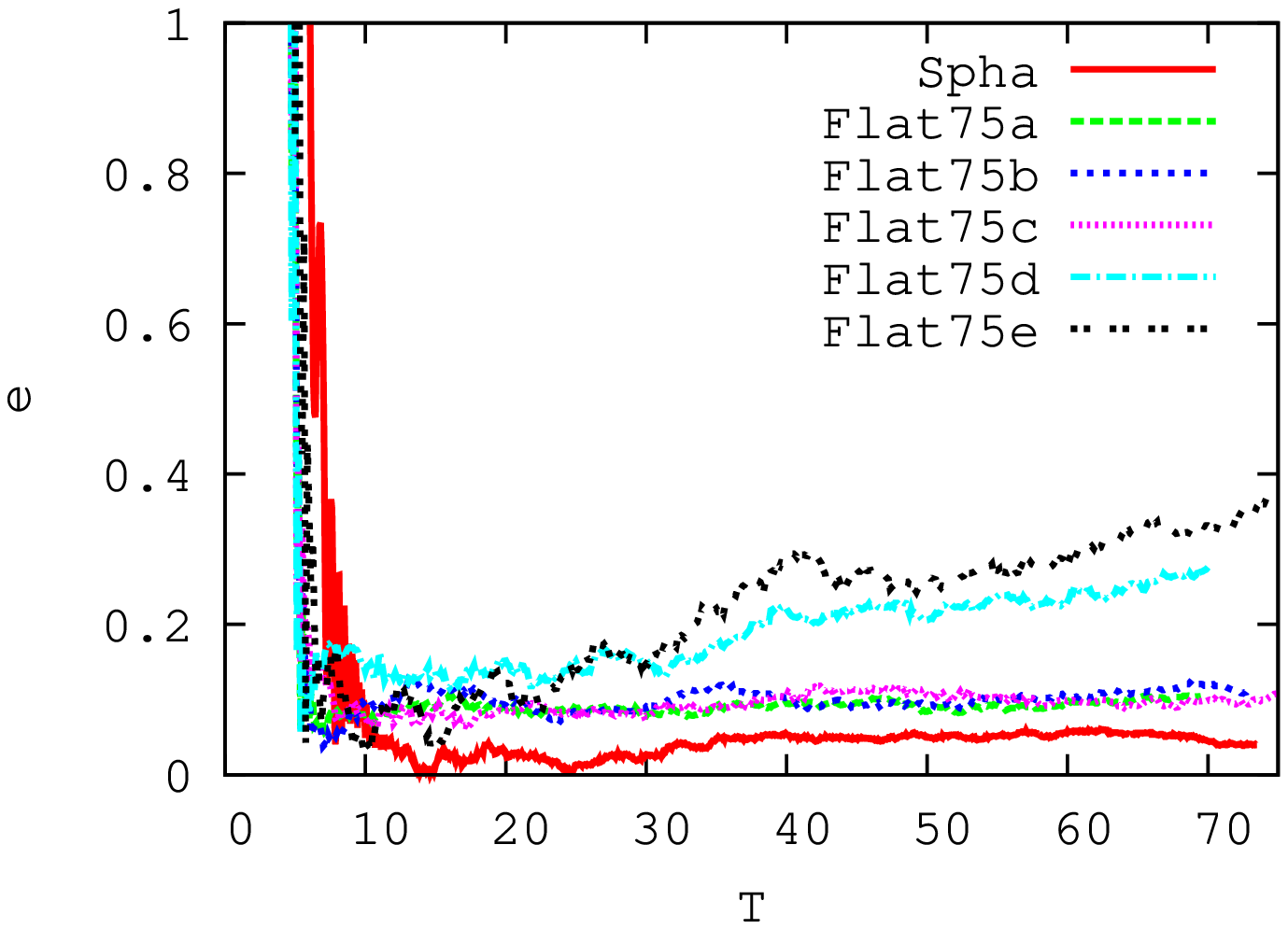}}
  }
\caption[]{
Evolution of the binary SMBH separation (top), semimajor axis (middle) and eccentricity (bottom) of the SMBH binary in $N$-body integrations for spherical and flattened models with $c/a = 0.75$  (see table \ref{TableA}). Time and $R$ are measured in model units.
} \label{flat4}
\end{figure}

Figure \ref{flat4} features the SMBH binary evolution for the flattest model we explored so far, $c=0.75$, and we confirm the both N-independence and the
rapid hardening rate seen in the $c=0.8$ runs. As is perhaps expected, the hardening rate is slightly enhanced compared to the $c=0.8$ model and the SMBHs reach an
even smaller separation. 


\subsection{Estimates for Relativistic Regime}

At the end of our simulations, the SMBH binary separation is still far larger than what is required for gravitational wave emission to dominate and guide the binaries on their way to coalescence. Earlier studies \citep{ber09,kh12} have shown that the estimated coalescence time obtained using constant hardening rate $s$ in the stellar dynamical regime and the formula of \citet{P64} for hardening in the gravitational wave dominated regime agree remarkably well with $T_{coal}$ obtained from simulations that follow the binary evolution until coalescence using post-Newtonian ($\mathcal{PN}$) terms in the equation of motion of the SMBH binary. 

So, we will use the \citet{P64} formalism to estimate the evolution of the SMBH binary beyond the Nbody regime~\citep[see also][]{kh12a}:

\begin{equation}
\frac{da}{dt} = \left(\frac{da}{dt}\right)_\mathrm{NB} + \Big\langle\frac{da}{dt}\Big\rangle_\mathrm{GW} = -s a^{2}(t) + \Big\langle\frac{da}{dt}\Big\rangle_\mathrm{GW} \label{ratea}
\end{equation}

The orbit-averaged expressions---including the lowest order 2.5 $\mathcal{PN}$ dissipative terms--- for 
the rates of change of a binary's semi-major axis, and eccentricity due to gravitational wave emission are
given by \citet{P64}:
\begin{mathletters}
\begin{eqnarray}
\Big\langle\frac{da}{dt}\Big\rangle_\mathrm{GW} &=& -\frac{64}{5}\frac{G^{3}M_{\bullet1}M_{\bullet2}(M_{\bullet1}+M_{\bullet2})}{a^{3}c^{5}(1-e^{2})^{7/2}}\times \nonumber \\
&&\left( 1+\frac{73}{24}e^{2}+\frac{37}{96}e^{4}\right),  \label{dadt}\\
\Big\langle\frac{de}{dt}\Big\rangle_\mathrm{GW}  &=& -\frac{304}{15}e\frac{G^{3}M_{\bullet1}M_{\bullet2}(M_{\bullet1}+M_{\bullet2})}{a^{4}c^{5}(1-e^{2})^{5/2}}
\times\nonumber\\
&&\left( 1+\frac{121}{304}e^{2}\right) .  \label{dedt}
\end{eqnarray}
\end{mathletters}

In order to compute the evolution in the gravitational radiation regime, we must adopt some physical length and mass. We use the observed mass of 
the SMBH and its influence radius for NGC4486A and compare it to the mass and influence radius of our model. One unit of length and mass in our model units equals $0.3$ kpc and $2.6 \times 10^9$ solar masses, respectively. The value of speed of light is $c = 1550$ and one time unit is equal to 1.4 Myr.

We next present an estimate of the coalescence time for the run Flat75a, which is a typical representation of the $c/a=0.75$ runs.  At T = 70 in model units, the semi-major axis $a$ is $5 \times 10^{-4}$ and the eccentricity $e$ is 0.1, which makes the estimated coalescence 2.4 Gyr. This coalescence is a few times longer than the coalescence times for radial galaxy mergers in  \cite{kh12}, but it is still well within a Hubble time. We speculate that the main reason behind this difference is that the eccentricity here is quite low. Indeed, we are currently exploring the eccentricity evolution in axisymmetric galaxy models to determine if low eccentricity is a generic feature of flattened equilibrium models.

\section{Summary \& Conclusions}\label{sec-concl}

We present the first results of SMBH binary evolution in a fully self-consistent equilibrium axisymmetric galaxy model. 
We find, in general, that SMBH binaries evolve faster and reach
smaller separations for even mildly flattened galaxy models. For $c/a=0.8$, we begin to see hints of $N$-independence for $N>500k$, and here the hardening rate is 25 times faster than 
in a spherical model. For our flattest model $c/a=0.75$, the system evolves to a gravitational radiation regime regardless of the particle resolution; for sufficiently flattened systems, we show that
the final parsec problem is not a problem. These models may best pertain to
pseudobulges and S0s; a recent example is the lenticular galaxy NGC 1277, which seems to host a new class of ultramassive black hole~\citep{vdb12}. Note, though, that the flattening in our models is very modest compared to these systems, where the axis ratios are often $c/a=0.6$ or smaller~\citep{Kormendy04}, and
we have also yet to include rotation. Both of these effects should act to increase the hardening rate. 

Contrary to previous work that consistently predict highly eccentric SMBH binaries as they pass to the gravitational radiation regime \citep{kh11,kh12,kh12a}, we
find that the SMBH binaries here all have quite low eccentricity, with the most eccentric orbit only having $e=0.1$. It is unclear whether this is
due to axisymmetry or to the fact that previous models were products of radial mergers, and therefore the system itself  -- both the SMBH and the 
particles in the model --  contained a large fraction highly radial orbits. More work is needed, both to examine the orbital content of our equilibrium models, 
and to explore the effect of the SMBH initial eccentricity, to unravel the eccentricity evolution of these SMBH binaries.

In this pilot study, all of our flattened galaxy models had the same density profile, central black hole mass, black hole mass ratio, and initial black hole orbital
eccentricity. Since the SMBH binary evolution will likely depend on the internal structure of the host galaxy and on details of the black hole
orbit, we are launching a systematic study of SMBH binary evolution in axisymmetric galaxy models of various cusp slopes, mass ratios, flattening and orbit type.

\acknowledgments

This work was conducted at the Advanced Computing Center 
for Research and Education at Vanderbilt University, Nashville, TN.  K. H.-B. acknowledges
support from NSF CAREER award AST-0847696, as well as the support from the Aspen Center for Physics. K.H.-B. also wishes to thank Daniel Sissom for assistance in this project.

\end{document}